\documentclass[twocolumn,prb,showpacs,superscriptaddress,amsmath,amssymb,natbib]{revtex4}

\usepackage{graphicx}
\usepackage[english,german]{babel}
\usepackage{amsmath}
\usepackage{amssymb}
\usepackage[latin1]{inputenc}

\begin{document}

\title{Characterization of local dynamics and mobilities in polymer melts - a simulation study}

\author{Diddo Diddens}
\email{d.diddens@uni-muenster.de}
\affiliation{Institut f\"ur physikalische Chemie, Westf\"alische Wilhelms-Universit\"at M\"unster, Corrensstrasse 28/30, 48149 M\"unster, Germany}
\affiliation{NRW Graduate School of Chemistry, Corrensstrasse 36, 48149 M\"unster, Germany}

\author{Martin Brodeck}
\affiliation{Institut f\"ur Festk\"orperforschung, Forschungszentrum J\"ulich GmbH, 52425 J\"ulich, Germany}

\author{Andreas Heuer}
\affiliation{Institut f\"ur physikalische Chemie, Westf\"alische Wilhelms-Universit\"at M\"unster, Corrensstrasse 28/30, 48149 M\"unster, Germany}
\affiliation{NRW Graduate School of Chemistry, Corrensstrasse 36, 48149 M\"unster, Germany}

\selectlanguage{english}

\date{\today}

\begin{abstract}
The local dynamical features of a PEO melt studied by MD simulations are compared 
to two model chain systems, namely the well-known Rouse model as well as the semiflexible 
chain model (SFCM) that additionally incorporates chain stiffness.
Apart from the analysis of rather general quantities such as the mean square displacement (MSD), 
we present a new statistical method to extract the local bead mobility from the simulation data 
on the basis of the Langevin equation, thus providing a complementary approach to the classical 
Rouse-mode analysis.
This allows us to check the validity of the Langevin equation and, as a consequence, the Rouse model.
Moreover, the new method has a broad range of applications for the analysis of the dynamics of more 
complex polymeric systems like comb-branched polymers or polymer blends.
\end{abstract}



\maketitle

\section{Introduction}

Characterizing the complex dynamics of a polymer chain in a melt
via simplifying models is an important problem in polymer physics.
For non-entangled chains the Rouse model \cite{Rouse,DoiEdwards}
can be regarded as the standard model. Within this model a polymer
chain is regarded to consist of $N$ harmonically bound beads, characterizing
the intra-molecular forces. Furthermore, the inter-molecular
interactions are described by stochastic forces, uncorrelated in
time and space. Considering the overdamped limit one can therefore
formulate the resulting Langevin equation. For bead $n$ of a bead-spring 
chain it is given by
\begin{equation}
  \label{Langevin_eq}
  \frac{d{\bf R}_n\left(t\right)}{dt}=-\frac{k}{\zeta}\left(2{\bf R}_n\left(t\right)-{\bf R}_{n+1}\left(t\right)-{\bf R}_{n-1}\left(t\right)\right)+\frac{1}{\zeta}{\bf F}_{n}\left(t\right),
\end{equation}
where $\zeta$ is the monomeric friction coefficient and
$k=3k_\mathrm{B}T/b^2$ is the entropic force constant of the
springs with the average squared length $b^2$. The amplitude of
the stochastic Brownian force is determined by the
fluctuation-dissipation theorem \cite{DoiEdwards}
\begin{equation}
  \langle F_{n,\alpha}\left(t\right)F_{m,\beta}\left(0\right)\rangle=2k_\mathrm{B}T\zeta\delta_{nm}\delta_{\alpha,\beta}\delta\left(t\right)\,,
\end{equation}
where $n$ and $m$ denote the monomer indices and $\alpha$ and
$\beta$ denote the three spatial directions.

An important observable frequently  discussed in the context of
simulations of polymeric systems and the elucidation of the
quality of the Rouse model is the mean square displacement (MSD)
\cite{PaulSmithRev2004,PaulChemPhys2002,BulacuvanderGiessenJCP2005,SteinhauserBlumenJCP2009}.
Naturally, the MSD of the monomers of a polymer chain shows
subdiffusive motion on a time scale below the longest relaxation
time $\tau_\mathrm{R}$ due to the chain connectivity.

Although in general reasonable agreement with the very simple
Rouse model and simulation data is observed, some specific
deviations are reported. First, from comparison of the MSD with
NSE experiments clear deviations from Gaussian behavior were
reported
\cite{PaulSmithRev2004,PaulSmithMamol2002,PaulSmithJCP2001}.
Second, rather than the predicted proportionality of
${\mathrm{MSD}}\propto{t^{\alpha}}$ with $\alpha=0.5$ for time
scales $\tau_\mathrm{R}/N^2 \le t \le \tau_\mathrm{R}$ simulations
of more realistic polymer chains yield
${\alpha}\approx{0.6}$
\cite{PaulSmithRev2004,PaulChemPhys2002,BulacuvanderGiessenJCP2005}. Third, memory
effects for the stochastic forces were observed
\cite{PaulSmithJCP2001,PaulPRL1998,PaulSmithRichterChemPhys2000},
giving rise to the formulation of more complicated approaches like
the generalized Langevin equation (GLE) and mode coupling theory
(MCT) \cite{SchweizerRev1997,GuenzaJCP1999,GuenzaPRL2001}. Forth,
the short-time monomer displacement, which reflects the motion on
a local scale, will additionally contain dynamical features
imposed by the detailed chemical structure which go beyond the predictions
of the Rouse model.

Alternatively, the quality of the Rouse model is probed by analysis
of the normal modes which in the ideal case display a specific 
dependence on the mode number when analyzing their amplitudes and 
their relaxation times \cite{Rouse,DoiEdwards}.
However, the deviations mentioned above are also observed in the
behavior of the higher-order Rouse modes 
\cite{PaulSmithRichterChemPhys2000,PaulSmithJCP2001,BulacuvanderGiessenJCP2005}.

The observables, reported so far, are either the normal modes themselves or,
in case of the MSD, can be expressed as a sum over the different
normal modes. The specific time-dependence of a mode, however, is
naturally determined by the intra- and inter-chain contributions
in eq. \ref{Langevin_eq}, the latter containing the complex
interaction with adjacent chains. As a consequence, a direct
identification of the intra-chain contributions is not possible
and, as a result, the MSD is  prone to all different non-idealities.

More generally, one faces the situation that in order to elucidate
the quality of the {\it local} Langevin equation one typically
analyzes {\it non-local} observables such as normal modes or their
sum. Here we present a new approach (pq-method) which allows us to
directly identify the local intra-molecular interactions in eq.
\ref{Langevin_eq}. This approach is applied to three different
model systems. First, we analyze an ideal Rouse chain in order to
verify our statistical approach. Second, we use MD simulations of
a chemically realistic PEO melt and compare the results with the
predictions from the Rouse model. Third, as a second model system
we consider the semiflexible chain model (SFCM)
\cite{WinklerEPL1999,WinklerJCP1994} which is supposed to be
superior to the Rouse model to describe the dynamics of chemically
realistic polymers.

After introduction of the pq-method we mention some simulation
details for all three systems (PEO, Rouse, SFCM). We start with the discussion of standard observables, namely
the MSD and some characteristics of forward-backward correlations. 
Then we apply the pq-method to the three different model
systems and characterize in detail the nature of the intra-chain interaction.
We conclude by indicating some interesting future
applications of the pq-method.

\section{pq-Method}

We start by defining the quantities
\begin{equation}
  {\bf p}_{n}\left(t,\Delta t\right)=\frac{{\bf R}_{n}\left(t+\Delta t\right)-{\bf R}_{n}\left(t\right)}{\Delta t}
  \label{p_def}
\end{equation}
and
\begin{equation}
  {\bf q}_{n}\left(t\right)=2{\bf R}_{n}\left(t\right)-{\bf R}_{n+1}\left(t\right)-{\bf
  R}_{n-1}\left(t\right).
  \label{q_def}
\end{equation}
Then eq. \ref{Langevin_eq} can be rewritten in the form
\begin{equation}
  \label{Langevin_eq_slope2}
  \lim_{\Delta t \rightarrow 0} {\bf p}_{n}\left(t,\Delta t\right)=-\frac{k}{\zeta}{\bf q}_{n}\left(t\right)+\frac{1}{\zeta}{\bf
  F}_{n}\left(t,\Delta t\right)\,.
\end{equation}
This is a linear relationship between the vectors ${\bf p}_n$ and
${\bf q}_n$ with a random component ${\bf F}_{n}$. The effective
bead mobility $k/\zeta$ can now easily be determined statistically
by performing a linear regression, {\it i.e.}
\begin{equation}
  A:= \frac{k}{\zeta}=\frac{\langle {\bf pq}\rangle-\langle {\bf p}\rangle\langle {\bf q}\rangle}{\langle {\bf q}^2\rangle-\langle {\bf q}\rangle^2}=\frac{\langle {\bf pq}\rangle}{\langle {\bf q}^2\rangle}\,.
  \label{slope_def}
\end{equation}
Strictly speaking one has to consider the limit of small $\Delta
t$ and can average over all inner monomers and all times (because
the dynamics is stationary).
Naturally, for a Rouse chain one would recover the effective bead mobility as it entered the calculation.

When applying the Rouse model to a chemically realistic polymer
chain such as PEO one must be more careful. As a key problem
eq. \ref{Langevin_eq} cannot be applied  because of ballistic
contributions for very small $\Delta t$ and non-universal local
contributions for somewhat longer $\Delta t$. This problem can be
circumvented as follows. We generalize eq.
\ref{Langevin_eq_slope2} to
\begin{equation}
  \label{Langevin_eq_slope}
 {\bf p}_{n}\left(t,\Delta t\right)=-\frac{k}{\zeta\left(\Delta t\right)}{\bf q}_{n}\left(t\right)+\frac{1}{\zeta}{\bf
  F}_{n}\left(t,\Delta t\right)\,.
\end{equation}
In this way we introduce a time-dependent friction coefficient
$\zeta(\Delta t)$, which for small $\Delta t$ approaches the bare
friction coefficient. The slope is determined in analogy to
eq. \ref{slope_def} by
\begin{equation}
  A(\Delta t)=\frac{\langle {\bf p}(\Delta t){\bf q}\rangle}{\langle {\bf q}^2\rangle}.
  \label{slope_def_dt}
\end{equation}

Strictly speaking the specific value of ${\bf
p}_{n}\left(t,\Delta t\right)$, reflecting the dynamics of the
n-th monomer during the {\it finite} time $\Delta t$, additionally contains
contributions beyond its direct neighbors. Exactly these
contributions will decrease the relative contribution of ${\bf q}_n(t)$
to the overall dynamics. This will
show up as a decrease of $k/\zeta(\Delta t)$ with increasing
$\Delta t$ (see below). However, this just reflects an intrinsic
problem of the Rouse model. The model is applicable only for some
finite time scale $\Delta t$  for which the Langevin equation in a
strict sense is no longer valid. By comparison with
$k/\zeta(\Delta t)$ of an ideal Rouse chain one can directly check
the applicability of the Rouse model to the PEO dynamics.

For a direct comparison of PEO with an ideal Rouse chain care must
be taken in the definition of eq. \ref{q_def} because Gaussian chain
properties have to be assured. Since adjacent chemical monomers are not
independent in their orientation, the distance of the two bond
vectors entering ${\bf q}$ have to be of the order of the Kuhn
length $b_\mathrm{K}$. In order to achieve this we rewrite
eq. \ref{q_def}
\begin{equation}
  {\bf q}_n\left(t\right)=2{\bf R}_{n'}\left(t\right)-{\bf R}_{n'+\Delta n'}\left(t\right)-{\bf R}_{n'-\Delta n'}\left(t\right)\,,
  \label{q_redef}
\end{equation}
where $n'$ numbers the chemical monomers and $\Delta n'$ has to be
chosen large enough that  Gaussian chain properties are assured to
a good approximation. The value of $\Delta n'$ can be estimated by
the characteristic ratio $C_\infty$ of the simulated chain by
making use of the identity
\begin{equation}
  C_\infty={\frac{b_\mathrm{K}}{b_0}}\approx{\left(\frac{\langle \left({\bf R}_{n'}-{\bf R}_{n'-\Delta n'}\right)^2\rangle}{\langle \left({\bf R}_{n'}-{\bf R}_{n'-1}\right)^2\rangle}\right)^\frac{1}{2}}\,,
  \label{delta_n_def}
\end{equation}
where $b_0$ is the average bond length between two chemical
monomers. For PEO, we obtain $\Delta n' = 4$. The model chain described
via the SFCM is treated in analogy to PEO. Since its bending
potential is optimized to reflect the structural properties of PEO
we again obtain $\Delta n' = 4$.

\section{Simulation Details}

For our analysis, MD simulation data of PEO from a previous study
have been used \cite{MaitraHeuerMCP2007}. A
system consisting of $16$ PEO chains with $48$ monomers each had
been simulated in an $NVT$ ensemble with the GROMACS simulation
package using the two-body effective polarizable force field
described in ref. \cite{BorodinSmithJPCB2003}. The temperature had
been maintained at $T=450\,\mathrm{K}$ by a Nos\'{e}-Hoover
thermostat.	

An idealized Rouse chain has been simulated via Brownian Dynamics
Simulations.  The ideal Rouse chain consists of $N=16$ monomers,
which corresponds to the number of effective beads of the PEO
chains with $N=48$ monomers and $C_\infty=3.2$ as determined from
the MD simulations.

Furthermore, Brownian Dynamics Simulations have been also
performed for the SFCM. In this model a Rouse chain is
supplemented by the additional potential
\begin{equation}
  U_\theta\left(\{{\bf R}_n \}\right) = k_\theta\sum_{n=2}^{N-1}\frac{\left({\bf R}_{n+1}-
                    {\bf R}_{n}\right)\left({\bf R}_{n}-{\bf R}_{n-1}\right)}{ | {\bf R}_{n+1}-{\bf R}_{n} || {\bf R}_{n}-{\bf R}_{n-1} |
                    }.
\end{equation}
Here the value of $k_\theta$ has been chosen such that the
characteristic ratio of the model chain matches that of PEO in the
MD simulations. Of course, the chains of the SFCM contained
the same number of monomers ($N=48$) as the PEO chains.

Using unit values for the friction coefficient $\zeta$, the
temperature $k_B T$ and the mean squared bond length $b^2$, a
time-step of $\Delta t=0.002$ turned out to be sufficiently small.

For the determination of $\tau_\mathrm{R}$, that will be used for
normalization in the following,  the autocorrelation function of
the longest Rouse mode was fitted via  \cite{DoiEdwards}
\begin{equation}
  \left<{\bf X}(0){\bf X}(t)\right>=\left<{\bf X}^2(0)\right>\exp(-t/\tau_\mathrm{R}).
  \label{Re2_decay}
\end{equation}
In the following calculations, the two outermost beads of the
Rouse chains and the five outermost monomers of the SFCM and the
PEO chains were ignored to exclude chain-end effects.

\section{Characterization of the Dynamics}

\begin{figure}
  \includegraphics[scale=0.3]{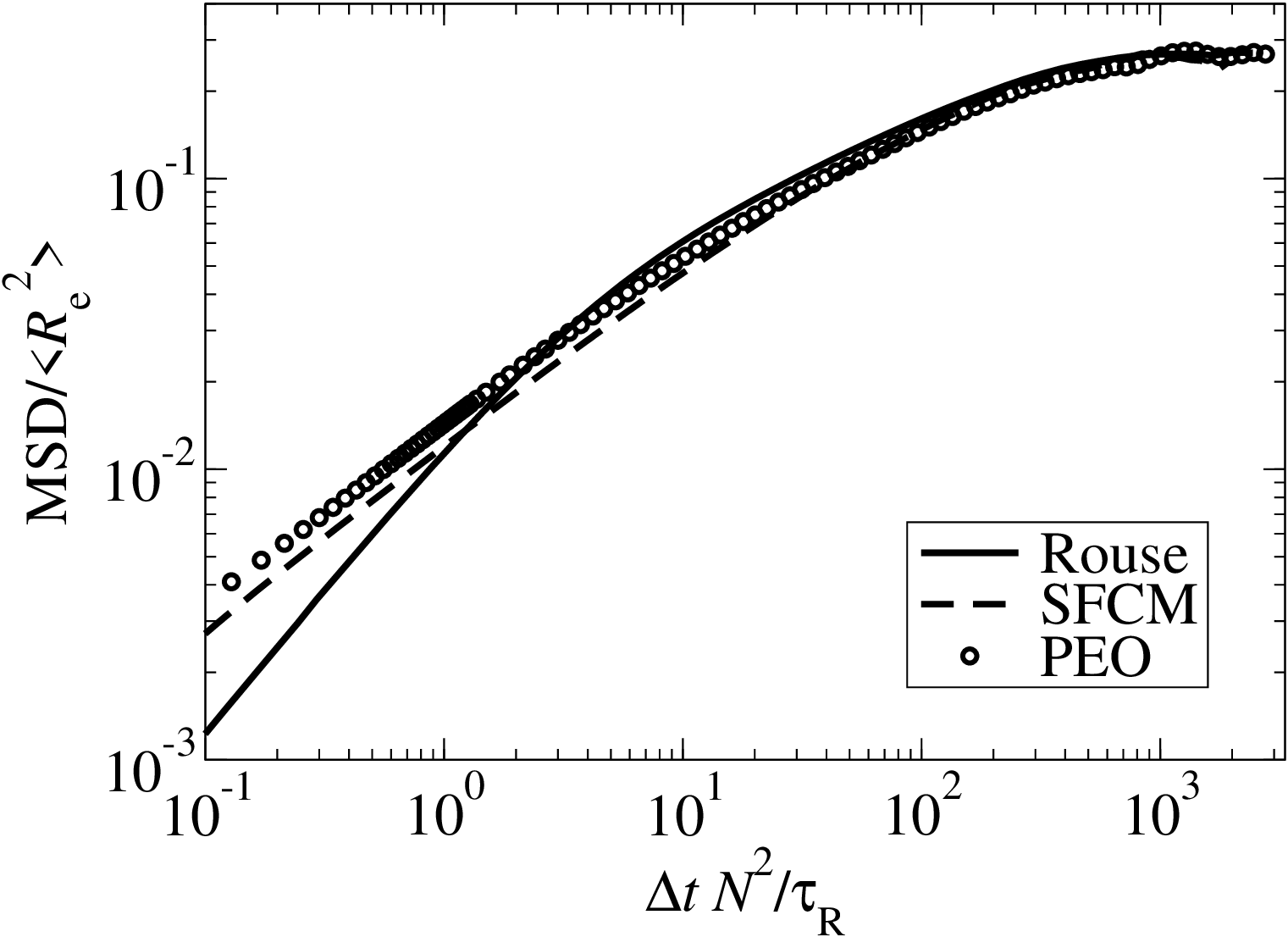}
  \caption{Mean square displacement (MSD) of the monomers in the center-of-mass frame for the Rouse chain, the SFCM and the PEO melt.}
  \label{cma_msd}
\end{figure}

We start by analyzing the MSD in the c.o.m.-frame (fig. \ref{cma_msd}). 
The scaling of the axes guarantees that the Rouse curve is independent of
the specific choice of the parameters $b$, $\zeta$, $T$ or $N$.
For later purposes we introduce $\Delta t_\mathrm{rel}=\Delta t N^2/\tau_\mathrm{R}$.
Naturally, the subdiffusive dynamics mainly reflects the chain connectivity. 
Evidently the SFCM reflects the PEO dynamics significantly better than the Rouse model, 
in particular for shorter times.
However, it is difficult to rationalize the nature of the residual deviations of 
the SFCM for shorter times.
A similar mismatch of the MSD-related quantity $S\left(q,t\right)$, {\it i.e.} the dynamic structure 
factor, for the Rouse model, the SFCM and an all-atom MD simulation was reported in ref. \cite{PaulSmithRichterChemPhys2000}.
By squaring eq. \ref{Langevin_eq_slope} and expressing the MSD as 
$\Delta t^2\langle{\bf p}^2\left(\Delta t\right)\rangle$ and using 
$\langle{\bf q}{\bf F}\rangle=0$ one obtains the contribution of the 
restoring and the effective stochastic forces to the overall MSD.
The latter is a summation of all forces not acting as restoring force 
within the correlator $\langle{\bf p}{\bf q}\rangle$ and therefore also 
containing the motion of more remote monomers 
and the inter-chain contributions.
We find for all time scales
that the contribution of the effective stochastic force is $3-10$ 
times higher (depending on $\Delta t$) than that of the restoring forces, making the MSD very sensitive 
to the special characteristics of the noise.

\begin{figure}
  \includegraphics[scale=0.3]{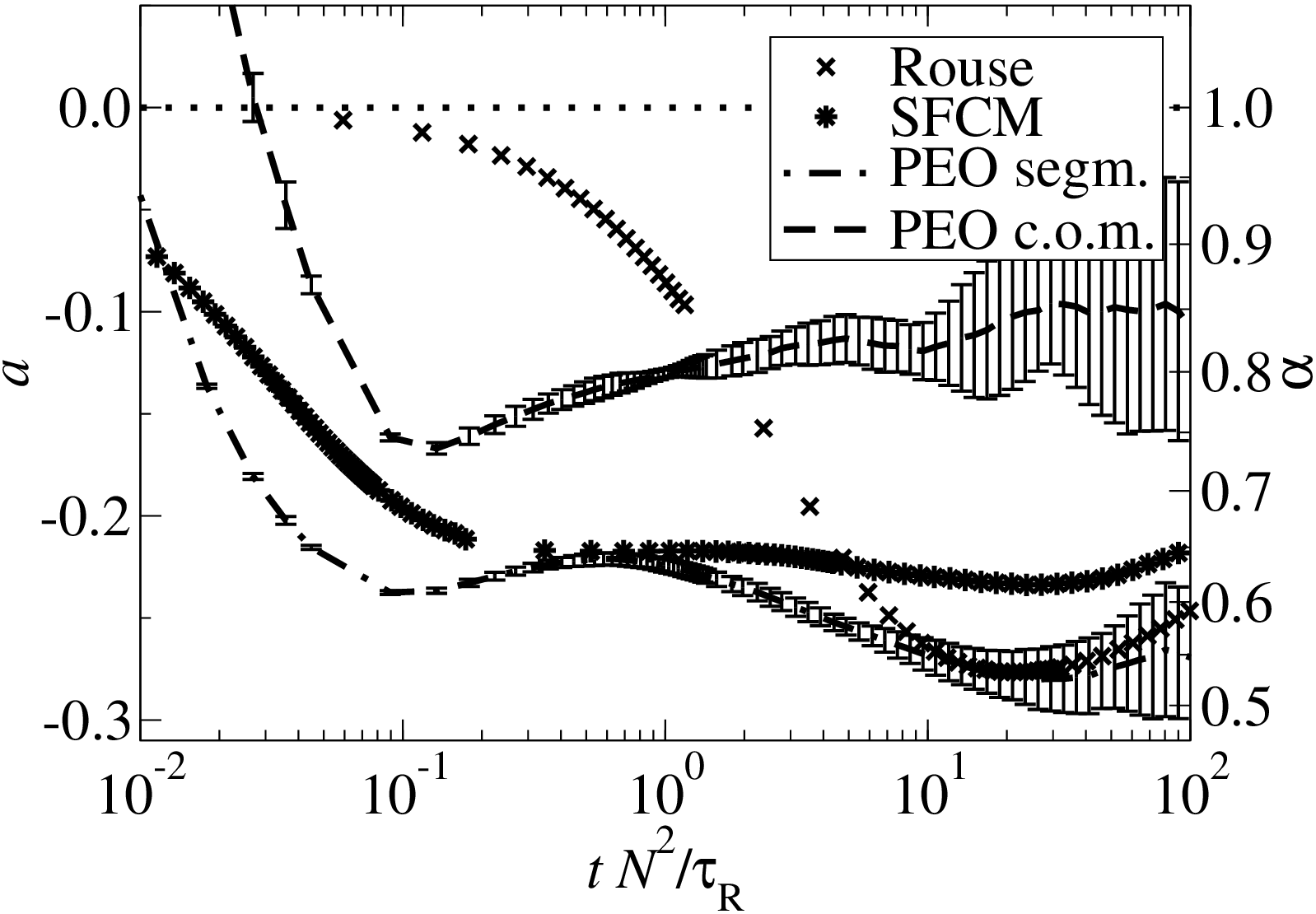}
  \caption{Correlation of two successive displacement vectors (left axis) and exponent of the MSD calculated by eqn \ref{alpha_eq} (right axis). The dotted line indicates the respective values of a freely diffusing particle.}
  \label{backjump_fig}
\end{figure}

A more detailed but related analysis is obtained by directly analyzing 
the properties of the subdiffusive dynamics.
For $\tau_\mathrm{R}/N^2 \le t \le \tau_\mathrm{R}$, the Rouse theory 
predicts a subdiffusive motion with a bead mean square displacement proportional to $t^{0.5}$.
In contrast to this, a proportionality for the monomer mean square displacement 
of ${\mathrm{MSD}\left(t\right)}\propto{t^\alpha}$ with ${\alpha}\approx{0.6}$ is 
observed in fig. \ref{cma_msd}, a value that is generally found in simulations of 
more realistic polymer models \cite{PaulSmithRev2004,PaulChemPhys2002,BulacuvanderGiessenJCP2005}.
This indicates that the backdriving forces in a real polymer melt are not only given 
by the chain connectivity, but also influenced by chain stiffness and intra- and intermolecular 
excluded volume effects.
As an investigation tool for this characteristic dynamics we define the following quantity
\begin{equation}
  a=\frac{\langle{\bf r}_1\left(t\right){\bf r}_1'\left(t\right)\rangle}{\langle{\bf r}_1^2\left(t\right)\rangle}\,,
  \label{a_def}
\end{equation}
where ${\bf r}_1\left(t\right)$ is the displacement of a given monomer 
during $t$ and ${\bf r}_1'\left(t\right)$ subsequent displacement during 
the next time interval of length $t$.
The MSD $\langle{\bf r}_2^2\left(2t\right)\rangle$ after $2t$ can also be 
written as the displacement after two successive steps with length $\langle{\bf r}_1^2\left(t\right)\rangle$ during $t$
\begin{equation}
  \label{r1r1pr2}
  \langle{\bf r}_2^2\left(2t\right)\rangle=2\langle{\bf r}_1^2\left(t\right)\rangle+2\langle{\bf r}_1\left(t\right){\bf r}_1'\left(t\right)\rangle\,,
\end{equation}
where the backdriving force is expressed by the second term on the rhs.
In cases where the MSD obeys a power law, {\it i.e.} $\langle{\bf r}_1^2\left(t\right)\rangle=\langle{\bf r}_1'^2\left(t\right)\rangle=ct^\alpha$ and $\langle{\bf r}_2^2\left(2t\right)\rangle=c\left(2t\right)^\alpha$, respectively, the exponent $\alpha$ can be related to the backward correlation $a$.
By dividing eq. \ref{r1r1pr2} by $\langle{\bf r}_1^2\left(t\right)\rangle$ and rearranging the expression one obtains
\begin{equation}
  \label{alpha_eq}
  \alpha=1+\log_2\left(1+a\right)\,.
\end{equation}
Figure \ref{backjump_fig} studies the backward correlation $a$ (left axis) and the 
exponent $\alpha$ (right axis) for the simulated Rouse chain, the SFCM and the PEO melt 
(the dotted line corresponds to a freely diffusing particle).
The dynamics of the Rouse model for very short times is essentially uncorrelated to 
its past, as the beads do not experience the connectivity constraint yet (fig. \ref{backjump_fig}).
For larger times one observes a long crossover to sublinear diffusion ending in a 
minimum with $\alpha=0.5$.
Note that the time window for this characteristic motion is rather short due to 
the short chain length ($N=16$).
In contrast to this, the local bending potential of the SFCM will immediately 
push back a monomer bending too far out of the chain curvature.
Thus, for the SFCM, the motion of a monomer is very early dominated by backjumps.
For larger time scales, $a$ remains nearly constant for the SFCM over several orders 
of magnitude with slight deviations from ideal behavior (${\alpha}\approx{0.6-0.65}$).

To a first approximation the PEO dynamics closely resembles the SFCM dynamics in agreement
with our observations from the MSD.
First, the dynamics of the PEO monomers crosses over from ballistic to backjump-dominated motion 
at ${\Delta t_\mathrm{rel}}\approx{0.1}$ ({\it i.e.} approximately $5\,\mathrm{ps}$, during which the 
monomers move on average $0.75$ times the chemical bond length $b_0$).
The minimum at ${\Delta t_\mathrm{rel}}\approx{0.1}$ can be identified as the confinement of the 
PEO monomers in a cage imposed by surrounding monomers in the first coordination sphere.
The same effect is also observed for low-molecular fluids \cite{DoliwaHeuerPRL1998}.
Second, for $\Delta t_\mathrm{rel}\geq 1$ the backward-correlations are again somewhat larger.
This can at least partly be related to the subdiffusive center-of-mass motion.
For real polymer melts, a slightly subdiffusive behavior (with ${\alpha}\approx{0.8}$) on a 
time scale shorter than the Rouse time was found in simulations 
\cite{PaulBinder1991,PaulPRL1998,PaulSmithRichterChemPhys2000,PaulSmithRev2004} and NSE experiments 
\cite{PaulSmithRichterChemPhys2000,GuenzaRichterJPCB2008} and is predicted by theoretical analyses 
\cite{GuenzaJCP1999,GuenzaPRL2001}.
This is also confirmed in fig. \ref{backjump_fig} where we also observe a slightly
subdiffusive behavior of the PEO chain's 
center-of-mass motion throughout the entire observation time larger than the ballistic regime.
The trend of the backward-correlation seems to become weaker with increasing time, 
however, we cannot make clear 
predictions within the error bars.
The additional backdriving motion acting on the PEO chains has been interpreted within the well 
known correlation hole picture \cite{deGennes}, in which regions of the chain that show correlated 
motion effectively repel each other for entropic reasons.
From theoretical and numerical calculations \cite{WittmerBaschnagelEPL2007} it was shown that the 
structure of a polymer chain in the melt can be viewed as a hierarchy of correlation holes throughout 
several length scales, where the repulsion is decreasing with increasing length scale.
Of course, for very long time scales standard diffusive behavior would be 
observed which is, however, beyond the time scales accessible by our simulations.

\section{Results of the pq-Method}

As motivated above the pq-method is supposed to yield a direct check of the intra-molecular interaction. 
In particular, the specific terms that cause the deviations of the MSDs average out in the quantity $A\left(\Delta t\right)$.

\begin{figure}
  \includegraphics[scale=0.3]{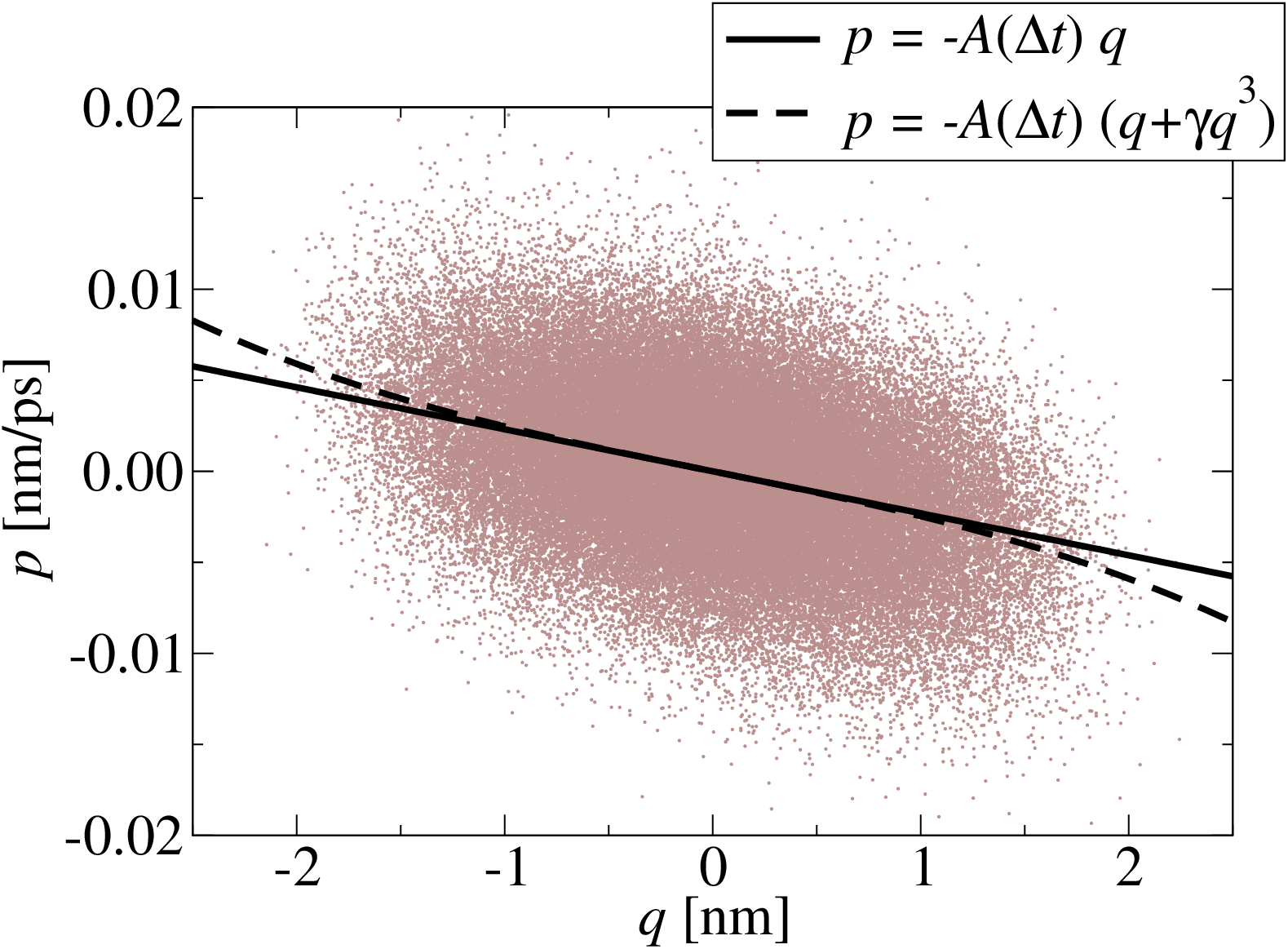}
  \caption{Scatterplot of the corresponding $p$- and $q$-values for the PEO chains with $\Delta n'=4$ and $\Delta t=84\,\mathrm{ps}$ as
  well as the regression line (eq. \ref{Langevin_eq_slope}) and an anharmonic third-order fit (eq. \ref{anharmonic_fit}).}
  \label{scatterplot}
\end{figure}

Figure \ref{scatterplot} shows a scatterplot of the $p$- and
$q$-values of the PEO melt with $\Delta t=84\,\mathrm{ps}$ as well
as the corresponding regression line with slope $A(\Delta
t)$ (solid line); see eq. \ref{slope_def_dt}.

To superimpose the curves obtained by eq. \ref{slope_def_dt}
we normalized $A(\Delta t)$ by the value
$k/\zeta=N^2/(\pi^2\tau_\mathrm{R})$ expected for the continuous
Rouse chain \cite{DoiEdwards}. As mentioned already above, $\Delta t$ was
normalized by $\tau_\mathrm{R}/N^2$, characterizing the local
relaxation time scale within the Rouse model.
In what follows the normalized time is denoted ${\Delta t_\mathrm{rel}}$.
\begin{figure}
  \includegraphics[scale=0.3]{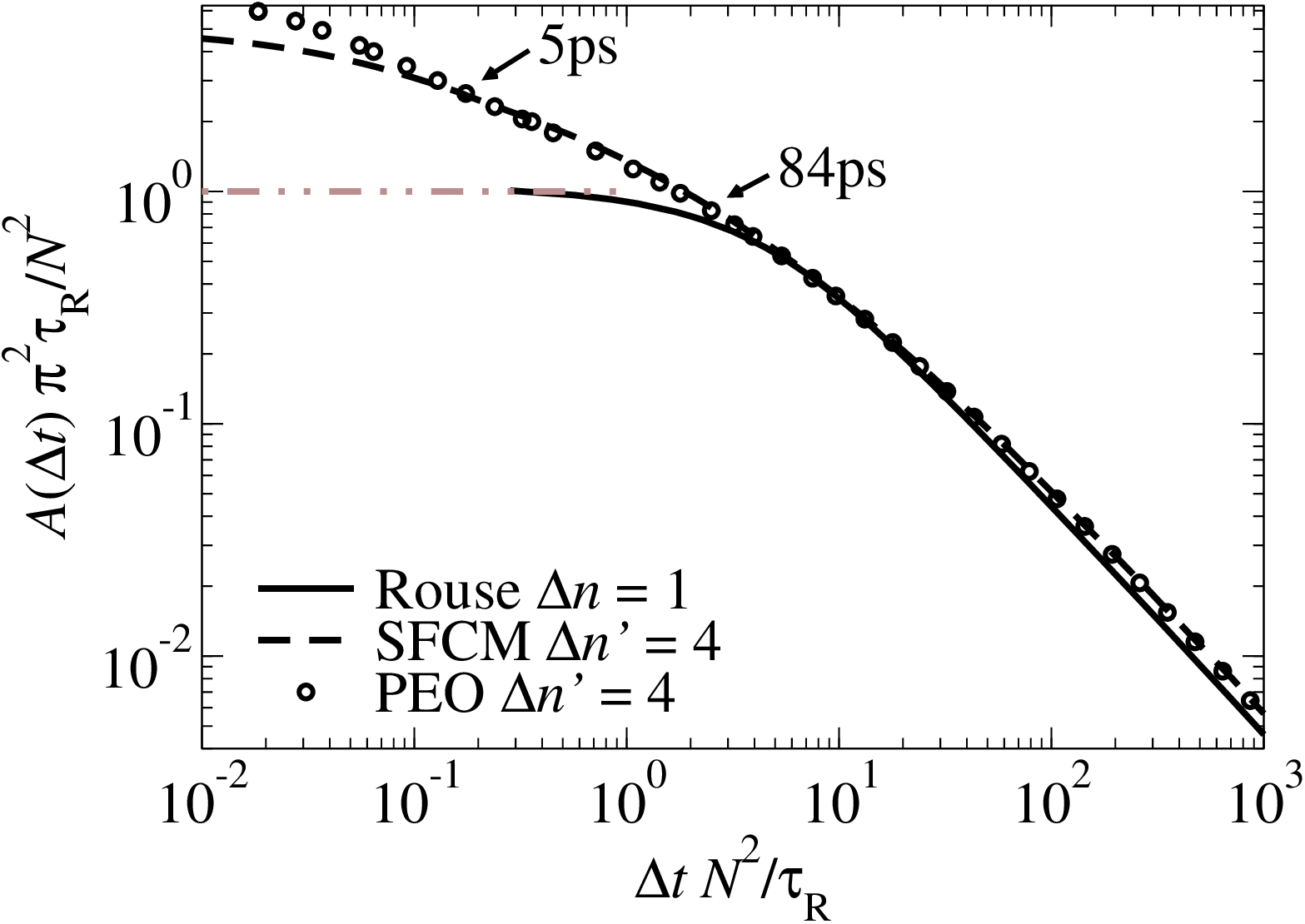}
  \caption{Normalized slope $A\left(\Delta t\right)$ as a function of normalized time
           for the three different systems analyzed in this work.
          }
  \label{A_delta_t}
\end{figure}

The normalized curves for $A_\mathrm{rel}(\Delta t_\mathrm{rel})$ for a Rouse
chain, the SFCM the PEO system are shown in fig.
\ref{A_delta_t}.
One observes that from about ${\Delta t_\mathrm{rel}}\approx{3}$ ({\it i.e.} ${\Delta
t}\approx{84\,\mathrm{ps}}$) the Rouse and the PEO curve agree
very well. At this time, a PEO monomer has moved
approximately twice the chemical bond length $b_0$, showing that
at this point the microscopic influence of the local potentials is
averaged out and the coarse-grained Rouse picture becomes
applicable. For shorter $\Delta t$ the curve of the Rouse chain
converges to the theoretical value of unity. Interestingly, around ${\Delta
t_\mathrm{rel}}\approx{3}$ the value of $A(\Delta t)$ for the
Rouse system has already reached this theoretical short-time limit.
One may conclude that at least for PEO the applicability of the
Rouse model holds without having to introduce renormalized values
for the friction coefficient.

In contrast to the Rouse model, for the SFCM and for PEO
$A\left(\Delta t\right)$ further increases with decreasing $\Delta
t$. This is a consequence of the chain stiffness which enhances
the effect of a local chain curvature. The PEO curve is in perfect
agreement with the SFCM from about ${\Delta
t_\mathrm{rel}}\approx{0.1\mathrm{ps}}$ (${\Delta
t}\approx{5\mathrm{ps}}$). Note that this is also the time scale for 
which the forward-backward coefficient $a$ becomes similar for both 
systems.  This result clearly shows that the
dynamical features of a PEO melt are basically captured by the
SFCM except for shortest times. The deviations for ${\Delta t} <
{5 \mathrm{ps}}$ are not due to inertia effects because the
velocity autocorrelation function in the MD simulation has already
fully decayed around $2\mathrm{ps}$. Thus, the specific properties
of the local PEO dynamics is only captured by the SFCM for
${\Delta t} > {5\mathrm{ps}}$.

\begin{figure}
  \includegraphics[scale=0.3]{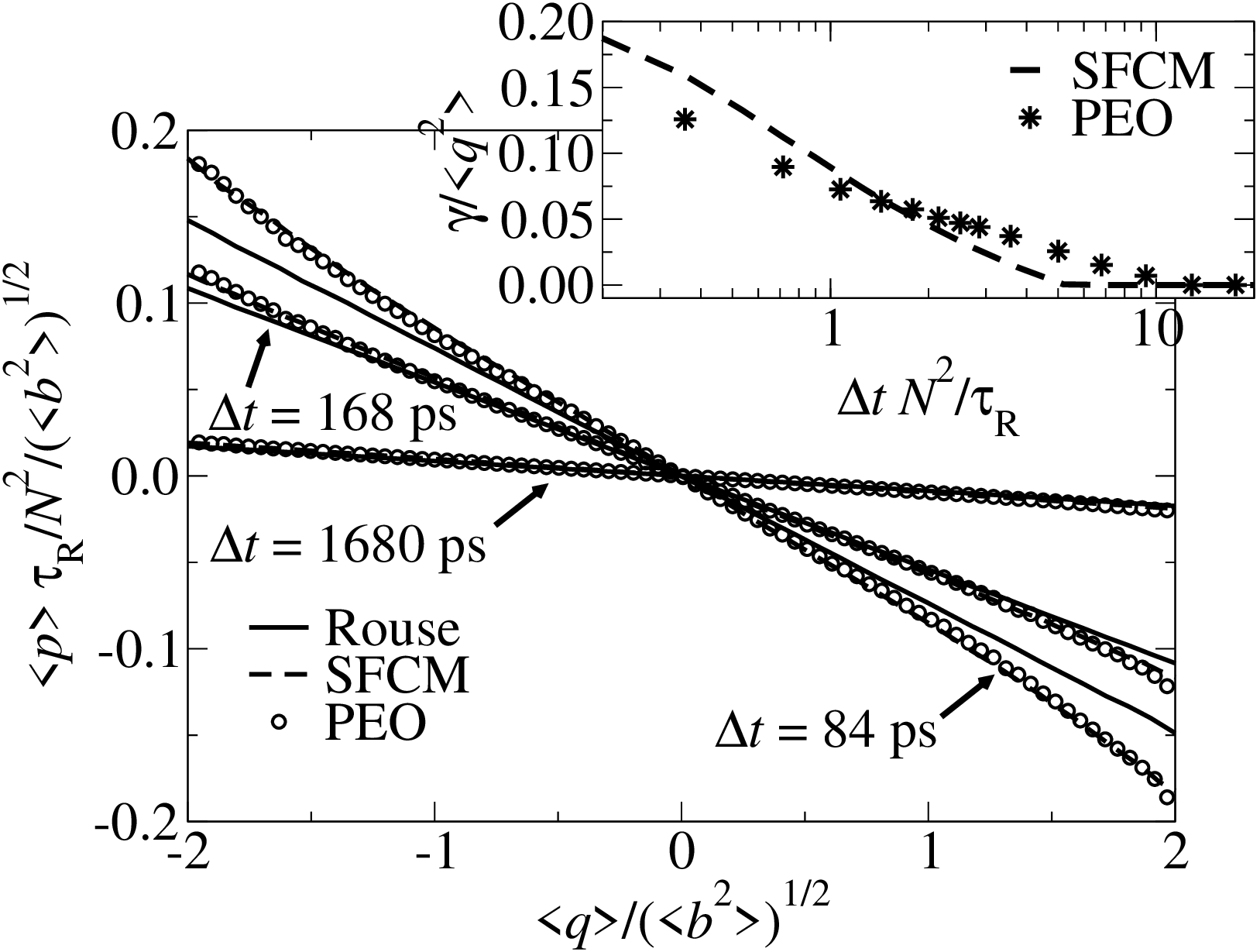}
  \caption{Normalized average $p$ and $q$ values from the histogram of the individual $q$ values for various $\Delta t$. 
           Inset: anharmonicity parameter $\gamma$ defined by eq. \ref{anharmonic_fit}.
          }
  \label{p_q_bin}
\end{figure}

According to the  Rouse model, the restoring force acting on a
given bead linearly depends on its elongation (eq.
\ref{Langevin_eq}), yielding the characteristic harmonic
potential. As can already be seen from fig. \ref{scatterplot},
an anharmonic fit including a third-order term is much more
appropriate to describe the data (dashed line). In the following,
we wish to examine the assumption of harmonic restoring forces for
the SFCM and PEO in more detail. The individual $q$-values were
assorted into a histogram, thus yielding the averages $\langle
q\rangle$ and $\langle p\rangle$ of the corresponding $p$-values
for each bin (fig. \ref{p_q_bin}).
For sufficiently small bead elongations the restoring force indeed
is linear. However, when going to larger $q$, this approximation
is no longer valid for PEO and the SFCM especially in the limit of
short $\Delta t$, and anharmonicities become noticeable. For the
SFCM and for PEO, the anharmonicities are similarly pronounced,
thus again highlighting that the SFCM effectively incorporates
many non-idealities of the PEO melt despite its simplicity. 
The inset in fig. \ref{p_q_bin} shows the
normalized anharmonicity parameter $\gamma$ defined by
\begin{equation}
  {\bf p}\left(\Delta t\right)=-A\left(\Delta t\right)\left({\bf q}+\gamma{\bf q}^3\right)
  \label{anharmonic_fit}
\end{equation}
when additionally considering third-order terms. The anharmonicity
expressed by $\gamma$ for both systems decays on the same
timescale. At ${\Delta t_\mathrm{rel}}\approx{3}$, where the
average  bead mobilities of PEO and the Rouse chain become
comparable (fig. \ref{A_delta_t}), $\gamma$ has nearly decayed
to zero. Although the SFCM and the PEO chains show an anharmonic
behavior at large elongations $q$, the linear dependence of the
restoring force as assumed in the Rouse model is valid to a good
approximation. Deviations become significant only on time scales
below the Rouse regime.
One might argue that the nonlinear dependence of $p$ on $q$ leads to
higher values of $A\left(\Delta t\right)$ for short $\Delta t$,
while for larger $\Delta t$ and vanishing anharmonicities the
curves of PEO and the SFCM converge to the Rouse curve (fig.
\ref{A_delta_t}). However, we find that the data points for large
$q$-values in figs. \ref{scatterplot} and \ref{p_q_bin}, where
higher-order terms become important, have a statistical weight
that is negligible compared to the $q$ values near zero.
Consistently, the overall value for $A\left(\Delta t\right)$ does
not change significantly when the large $q$-values are excluded
from the calculation.

\section{Conclusions and Outlook}

In this article, we presented the pq-method to extract the effective bead mobility from simulation data of polymeric systems. From 
a conceptual point of the view the main advantage is the direct accessibility of the intra-chain interaction effects. 
In contrast to this, the analysis of, {\it e.g.}, the MSD, immediately also involves possibly complex inter-chain contributions. 
In this way it became possible to show that the SFCM is a perfect model for PEO already for relatively short time scales.
Furthermore we could show that the dynamics of the discrete Rouse model is still characterized by the bare friction coefficient 
at the shortest relevant time scale of $84 \mathrm{ps}$.

However, also from a practical perspective the pq-method has a broad range of applications. 
For example it is possible to determine the friction coefficient of the $n$-th monomer. This may be of importance if
the friction coefficients are heterogeneously distributed. Examples are polymer mixtures ({\it e.g.} PEO/PMMA) where the local structure of the
slow component may determine the local mobility of the fast component. More generally, a detailed characterization of local dynamic
properties becomes accessible. In contrast, observables such as the MSD already contain in an uncontrolled  way dynamic contributions from
adjacent monomers or even complete Rouse modes, thus invalidating a strictly local analysis. 
Furthermore, the pq-method can be directly applied to the analysis of topologically more complex systems ({\it e.g.} branched polymers) where
the local friction coefficient may be directly calculated, {\it e.g.}, in dependence of the distance to the main chain.

\begin{acknowledgements}
D.D. and A.H. would like to thank the Sonderforschungsbereich 458 and the 
NRW Graduate School of Chemistry for financial support as well as 
Wolfgang Paul and Michael Vogel for helpful discussions.
M.B. would like to thank Reiner Zorn for helpful discussions.
\end{acknowledgements}

\bibliography{literatur}

\end{document}